 \documentstyle[12pt]{article}

\font\twelve=cmbx10 at 15pt
\font\ten=cmbx10 at 12pt

\def\upa{\uparrow}
\def\downa{\downarrow}
\def\cit#1{$^{[#1]}$}
\def\pr{Phys.\ Rev.\ }
\def\prl{Phys.\ Rev.\ Lett.\ }
\def\pl{Phys.\ Lett.\ }
\def\etal{{\it et al}.}

\begin{document}

\begin{titlepage}

\begin{center}

{\ten Centre de Physique Th\'eorique - CNRS - Luminy, Case 907}

{\ten F-13288 Marseille Cedex 9 - France }

{\ten Unit\'e Propre de Recherche 7061}

\vspace{1 cm}

{\twelve  EXPERIMENTAL EVIDENCE FOR SIMPLE
RELATIONS BETWEEN UNPOLARIZED AND
POLARIZED PARTON DISTRIBUTIONS}

\vspace{0.3 cm}

{\bf C. BOURRELY and
J. SOFFER}

\vspace{1.5 cm}

{\bf Abstract}

\end{center}

The Pauli exclusion principle is advocated for constructing the proton
and neutron deep inelastic structure functions in terms of Fermi-Dirac
distributions that we parametrize with very few parameters. It allows
a fair description of the recent NMC data on $F^p_2(x,Q^2)$ and
$F^n_2(x,Q^2)$ at $Q^2=4 GeV^2$, as well as the CCFR neutrino data at
$Q^2=3$ and $5 GeV^2$. We also make some reasonable and simple
assumptions to relate unpolarized and polarized quark parton
distributions and we obtain, with no additional free parameters, the
spin dependent structure functions $xg^p_1(x,Q^2)$ and
$xg^n_1(x,Q^2)$. Using the correct $Q^2$ evolution, we have checked
that they are in excellent agreement with the very recent SMC proton
data at $Q^2=10 GeV^2$ and the SLAC neutron data at $Q^2=2 GeV^2$.

\vspace{2 cm}



\noindent Number of figures : 6


\noindent May 1994

\noindent CPT-94/P.3032

\bigskip

\noindent anonymous ftp or gopher: cpt.univ-mrs.fr

\end{titlepage}

Many years ago Feynman and Field made the conjecture\cit{1} that the
quark  sea in the proton may not be flavor symmetric, more precisely
$\bar d>\bar u$, as a consequence of Pauli principle which favors $d\bar
d$ pairs with respect to $u\bar u$ pairs because of the presence of two
valence $u$ quarks and only one valence $d$ quark in the proton. This
idea was confirmed by the results of the NMC experiment\cit{2} on the
measurement of proton and neutron  unpolarized structure functions,
$F_2(x)$. It yields a fair evidence for a defect in the Gottfried sum
rule\cit{3} and one finds

\begin{equation}
I_G = \int^1_0 \frac{dx}{x} [ F^p_2(x)-F^n_2(x)] = 0.235\pm0.026
\label{1}
\end{equation}

\noindent instead of the value 1/3 predicted with a flavor symmetric
sea, since we have in fact

\begin{equation}
I_G = \frac{1}{3} (u+\bar u - d -\bar d) = \frac{1}{3} + \frac{2}{3}
(\bar u - \bar d). \label{2}
\end{equation}

\noindent A crucial role of Pauli principle may also be advocated to
explain the well known dominance of $u$ over $d$ quarks at high
$x$,\cit{4} which explains the rapid decrease of the ratio $F^n_2(x)
/ F^p_2(x)$ in this region. Let us denote by $q^\upa (q^\downa)$, $u$ or
$d$ quarks with helicity parallel (antiparallel) to the proton helicity.
The double helicity asymmetry measured in polarized muon (electron) -
polarized proton deep inelastic scattering allows the determination of
the quantity $A^p_1(x)$ which increases towards one for high
$x$,\cit{5,6} suggesting that in this region $u^\upa$ dominates over
$u^\downa$, {\it a fortiori} dominates over $d^\upa$ and $d^\downa$,
and we will see now, how it is possible to make these considerations
more quantitative. Indeed at $Q^2=0$ the first moments of the valence
quarks are related to the values of the axial couplings

\begin{equation}
u^\upa_{\rm val} = 1+F, \quad u^\downa_{\rm val}=1-F, \quad d^\upa_{\rm
val} = \frac{1+F-D}{2}, \quad d^\downa_{\rm val} = \frac{1-F+D}{2},
\label{3}
\end{equation}

\noindent so by taking $F=1/2$ and $D=3/4$ (rather near to the quoted
values\cit{7} $0.461\pm0.014$ and $0.798\pm0.013$) one has $u^\upa_{\rm
val} =3/2$ and $u^\downa_{\rm val} = 1/2$ which is at the center of the
rather narrow range $(d^\upa_{\rm val},d^\downa_{\rm val}) = (3/8,
5/8)$. The abundance of each of these four valence quark species,
denoted by $p_{\rm val}$, is given by eq.~(\ref{3}) and we assume that
the distributions at high $Q^2$ ``keep a memory'' of the properties of
the valence quarks, which is reasonable since for $x>0.2$ the sea is
rather small. So we may write for the parton distributions

\begin{equation}
p(x) = F(x, p_{\rm val}) \label{4} \end{equation}

\noindent where $F$ is an increasing function of $p_{\rm val}$. The fact
that the dominant distribution at high $x$ is just the one corresponding
to the highest value of $p_{\rm val}$, gives the correlation
{\it abundance -
shape} suggested by Pauli principle, so we expect broader shapes for
more abundant partons. If $F(x, p_{\rm val})$ is a smooth function of
$p_{\rm val}$, its value at the center of a narrow range is given, to a
good approximation, by half the sum of the values at the extrema, which
then implies\cit{8}

\begin{equation}
u^\downa_{\rm val}(x) = \frac{1}{2} d_{\rm val}(x) . \label{5}
\end{equation}

\noindent This leads to

\begin{equation}
\Delta u_{\rm val}(x) \equiv u^\upa_{\rm val}(x) - u^\downa_{\rm val}(x) =
u_{\rm val}(x) - d_{\rm val}(x)\label{6}
\end{equation}
and, in order to generalize this relation to the whole $u$ quark
distribution, we assume that eq.~(\ref{6}) should also hold for quark sea
and antiquark distributions, so we have

\begin{equation}
\Delta u_{sea}(x) = \Delta\bar u(x)=\bar u(x)-\bar d(x)\ .\label{7}
\end{equation}
Moreover as a natural consequence of eq.~(\ref{3}), we will
assume\footnote{It is amusing to remark that with the values of $F$ and
$D$ quoted above, we have in fact $\Delta d_{\rm val}(x)=-\frac{1}{3}
d_{\rm val}(x)$ which coincides with the so-called conservative $SU(6)$
model\cit{9}!}
\begin{equation}
\Delta d_{\rm val}(x) = (F-D)d_{\rm val}(x)\ .\label{8}
\end{equation}
Finally we will suppose that the $d$ sea quarks (and antiquarks) are not
polarized i.e.
\begin{equation}
\Delta d_{\rm sea}(x) = \Delta \bar d(x)=0\label{9}
\end{equation}
and similarly for the strange quarks
\begin{equation}
\Delta s(x) = \Delta \bar s(x)=0\ .\label{10}
\end{equation}
Clearly the above simple relations (6)-(10) are enough for fixing the
determination of the spin dependent structure functions
$xg^{p,n}_1(x,Q^2)$, in terms of the spin average quark parton
distributions. We now proceed to present our approach for constructing
the nucleon structure functions $F^{p,n}_2(x,Q^2)$, $xF^{\nu
N}_3(x,Q^2)$, etc... in terms of Fermi-Dirac distributions which is
motivated by the importance of the Pauli exclusion principle, as we
stressed above. A first attempt for such a construction was made in
ref.~[10], but here, as we shall see, our method is slightly
different and leads to significant improvements. Let us consider $u$
quarks and antiquarks only, and let us assume that at fixed $Q^2,
u^{\upa}_{\rm val}(x)$,
$u^{\downa}_{\rm val}(x)$, $\bar u^{\upa}(x)$ and $\bar u^{\downa}(x)$
are expressed in terms of Fermi-Dirac distributions, in the scaling
variable $x$, of the form
\begin{equation}
xp(x)=a_px^{b_p}/(exp((x-\tilde x(p))/\bar x)+1)\ .\label{11}
\end{equation}
Here $\tilde x(p)$ plays the role of the "thermodynamical potential"
for the fermionic parton $p$ and $\bar x$ is the "temperature" which is
a universal constant. Since valence quarks and sea quarks have very
different $x$ dependences, we expect $0<b_p<1$ for
$u^{\upa,\downa}_{\rm val}(x)$ and $b_p<0$ for $\bar
u^{\upa,\downa}(x)$. Moreover $\tilde x(p)$ is a constant for
$u^{\upa,\downa}_{\rm val}(x)$, whereas for $\bar u^{\upa,\downa}(x)$,
it has a smooth $x$ dependence. This might reflects, the fact that
parton distributions contain two phases, a gas contributing to the non
singlet part with a constant potential and a liquid, which prevails at
low $x$, contributing to the singlet part with a potential  slowly
varying in $x$, that we take linear in $\sqrt x$. In addition, in a
statistical model of the nucleon\cit{11}, we expect quarks and
antiquarks to have opposite  potentials, consequently
the gluon, which produces $q\bar q$ pairs, will have a zero potential.
Moreover since in the process $G\to q_{\rm sea} + \bar q$, $q_{\rm
sea}$ and $\bar q$ have opposite helicities, we expect the potentials
for $u^{\upa}_{\rm sea}$ (or $\bar u^{\upa}$) and $\bar u^{\downa}$ (or
$u^{\downa}_{\rm sea}$) to be opposite. So we take
\begin{equation}
\tilde x(\bar u^{\upa})=-\tilde x(\bar u^{\downa})=x_0+x_1\sqrt x\
.\label{12}
\end{equation}
The $d$ quarks and antiquarks are obtained by using eqs.~(\ref{5}) and
(\ref{7}) and concerning the strange quarks, we take in accordance with
the data \cit{12}
\begin{equation}
s(x)=\bar s(x)=\frac{\bar u(x)+\bar d(x)}{4}\ .\label{13}
\end{equation}
Finally for the gluon distribution, for the sake of consistency, one
should assume a Bose-Einstein expression given by
\begin{equation}
xG(x)=\frac{a_Gx^{b_G}}{e^{x/\bar x}-1}\label{13}
\end{equation}
with the same temperature $\bar x$ and a vanishing potential, as we
discussed above.

To determine our parameters we have used the most recent NMC
data\cit{2} on $F_2^p(x)$ and $F_2^n(x)$ at $Q^2=4 GeV^2$ together with
the most accurate neutrino data from CCFR\cit{12,13} on $xF^{\nu
N}_3(x)$ and the antiquark distribution $x\bar q(x)$ \cit{12}.

The universal temperature is found to be
\begin{equation}
\bar x=0.120\label{15}
\end{equation}
and for valence quarks we get the {\it three} free parameters
\begin{equation}
b(u^{\upa}_{\rm val})=1/2b(u^{\downa}_{\rm val})= 0.399,\ \tilde
x(u^{\upa}_{\rm val})=0.502,\ \tilde
x(u^{\downa}_{\rm val})=0.163 .\label{16}
\end{equation}
This relation between the $b's$ is imposed by the small $x$ behavior
of $xF_3^{\nu N}(x)$,
$a^{\upa}$ and $a^{\downa}$ are not free parameters, but two
normalization constants which are fixed from the obvious requierements
to have the correct number of valence quarks  in
the proton.  As we
noticed before
$u^{\upa}_{\rm val}(x)$ dominates, so it is not surprising to find that
it has a larger potential than $u^{\downa}_{\rm val}(x)$ \footnote{In a
statistical model of the nucleon\cit{11}, the potentials associated
with $u$ and $d$ quarks are taken in the ratio $2^{1/3}$ which is a much
smaller than the value of $\tilde x(u^{\upa}_{\rm val})/\tilde
x(u^{\downa}_{\rm val})\sim 3$ we have found.}.

For antiquarks we have {\it four} additional free parameters
\begin{equation}
\bar b=-0.358,\ \bar a^{\upa}=0.024,\ x_0=0.215\ \hbox{and}\
x_1=-0.388\ \hbox{for}\ \bar u^{\upa}.\label{17}
\end{equation}
$\bar b$ is the same for $\bar u^{\upa}$ and $\bar u^{\downa}$.

When $x\to 0$, from Pomeron universality, one expects $x\bar u(x)=x\bar
d(x)\not = 0$, so $\bar a^{\downa}$ is determined by this constraint.

We show the results of our fit for $F_2^p(x)-F_2^n(x)$ and
$F_2^n(x)/F_2^p(x)$ by the solid lines in Figs.~1a and b and for
$xF_3^{\nu N}(x)$ and $x\bar q(x)$ in Figs.~2a and b. The accuracy of
these neutrino data gives strong constraints on both valence and sea
quark distributions. The description of the data is very satisfactory
taking into account the fact that we only have {\it eight} free
parameters and this certainly speaks for Fermi-Dirac distributions. Note
that we find
$I_G=0.232$ in beautiful agreement with eq.~(\ref{1}). The steady rise
of $x\bar q(x)$ at small $x$ leads to a rise of $F^p_2$ which is
consistent with the first results from Hera\cit{14}. For the
fraction of the total momentum carried by quarks and antiquarks we find
\begin{eqnarray}
&\int^{1}_{0}xu(x)dx=0.298,\ \int^{1}_{0}xd(x)dx=0.115,\\\nonumber
&\int^{1}_{0}x[\bar u(x)+\bar d(x)+s(x)+\bar s(x)]dx=0.147.\label{18}
\end{eqnarray}
Concerning the gluon distribution, we find $a_G=14.536$ and $b_G=0.912$
and $xG(x)$ is fairly consistent with some preliminary indirect
experimental determination from direct photon production\cit{15},
from neutrino deep inelastic scattering \cit{16} at $Q^2=5 GeV^2$ and
at high $Q^2$ and smaller $x$ from NMC\cit{17}.

Let us now turn to the polarized structure functions
$xg_1^{p,n}(x,Q^2)$ which will allow to test our simple relations
(\ref{6})-(\ref{10}). Since the proton data from EMC\cit{5} and
SMC\cit{6} are at $Q^2=10 GeV^2$ and the neutron data from
SLAC\cit{18}\footnote{There is also some SMC data\cit{19} on the
polarized structure function $xg^d_1$ on deuterium.} is at
$Q^2=2GeV^2$, we have to consider the $Q^2$ evolution in order to use
our parton distributions determined at $Q^2=4GeV^2$. For this purpose
we have used a numerical solution\cit{20} of the Altarelli-Parisi
equations\cit{21} which lead to relatively small corrections in the
$Q^2$ range we are dealing with. In Figs.~1a, b the dashed lines are
the theoretical predictions at $Q^2=10 GeV^2$. As expected we see that
for low $x$, $F_2^p(x)-F_2^n(x)$ increases with $Q^2$ whereas it
decreases with $Q^2$ for high $x$, leaving the integral unchanged. The
$Q^2$ dependence of the ratio $F_2^n(x)/F_2^p(x)$ has the right trend
although probably a bit too weak compared to experimental observation
which has been attributed to different higher twist effects for proton
and neutron\cit{22}.

At this stage we would like to examine the consequence, of our simple
relations (\ref{6})-(\ref{10}). If the $d$ (valence and sea) quarks
were unpolarized, eqs.~(\ref{6}) and (\ref{7}) allow to relate the
contribution of $u$ quarks to $xg^p_1(x)$, to the contributions of $u$
and $d$ to $F_2^p(x)-F_2^n(x)$ i.e.
\begin{equation}
xg^p_1(x)\vert_u=\frac{2}{3}\left(F^p_2(x)-F^n_2(x)\right)\vert_{u+d}\
.\label{19}
\end{equation}
First, by comparing Fig.~1a and Fig.~3 \footnote{The vertical scales
have been chosen in such a way that one absorbs the factor $2/3$ by
superimposing the two figures.} one sees very clearly, the similarity of
the two sets of data points\footnote{This was first noticed in
ref.~[8] but, with more accurate data, it becomes now very convincing.}.

Second, on Fig.~3 the dashed line represents simply the dashed line of
Fig.~1a multiplied by a factor $2/3$, whereas the dotted line
corresponds to the case of a flavor symmetric sea, i.e. $\bar u(x)=\bar
d(x)$, or, in other words, to a zero polarization of the $u$ sea
quarks. This shows why we strongly suspect that the defect in the
Gottfried sum rule and the defect in the proton
Ellis-Jaffe sum rule\cit{23} are closely related. We still think it has
nothing to do with the polarization of the strange quarks, that we took
to be zero (see eq.~(\ref{10})), which is supported by reasonable
phenomenological arguments\cit{24}. Moreover, in this approach, the
strange quarks do not even participate, because they cancel in the
difference $F_2^p(x)-F_2^n(x)$. Finally, if one takes into account the
polarization of the $d$ valence quarks by using eq.~(\ref{8}), we get
the solid line in Fig.~3 which improves the agreement with the data. In
fact we found for $Q^2=10 GeV^2$
\begin{equation}
\int_{0.003}^{0.7} g^p_1(x)dx=0.136\label{20}
\end{equation}
in beautiful agreement with eq.~(\ref{4}) of ref.~[6].

Concerning the neutron polarized structure function $xg^n_1(x)$ we show
in Fig.~4 a comparison of the SLAC data\cit{18} at $Q^2=2GeV^2$ with
our theoretical calculations. The dashed line corresponds to the case
where $d$ quarks are assumed to be unpolarized and it clearly disagrees
with the data. However by including the $d$ valence quark polarization
according to eq.~(\ref{8}), we obtain the solid line in perfect
agreement with the data and we find for $Q^2=2GeV^2$
\begin{equation}
\int_{0}^{1} g^n_1(x)dx=-0.017\ .\label{21}
\end{equation}

To summarize we have given an accurate description of deep inelastic
scattering data at low $Q^2$ in terms of Fermi-Dirac distributions
parametrized with only eight free parameters for quarks and antiquarks.
Although we have some understanding of their meaning, much remains to be
done for a more fundamental theoretical interpretation, in terms of new
information for the nucleon structure. We have proposed a set of simple
relations between unpolarized and polarized quark (antiquark)
distributions for which, so far, there is a striking experimental
evidence. Of course our approach has to be further tested with more
accurate deep inelastic scattering data and in particular the important
issue of the validity of the Bjorken sum rule\cit{25}. Polarized proton
collisions at high energies will also provide independent tests which
will be most welcome in the near future\cit{26}.

\section*{Acknowledgments}

We thank F. Buccella for interesting discussions at the earlier stage
of this work.
We are very thankful to J. Kwiecinski for providing us his computer
programme. One of us (J.S.) is gratefull for kind hospitality at the
Theory Division (CERN) where part of this work was done.

\newpage

\section*{Figure Captions}

\begin{itemize}
\item[Fig.1a] The difference $F^p_2(x)-F^n_2(x)$ at $Q^2=4 GeV^2$
versus $x$. Data are from ref.[2] and the solid line is the result of
our fit. The dashed line is the theoretical result after evolution at
$Q^2=10 GeV^2$.

\item[Fig.1b] The ratio $F^n_2(x)/F^p_2(x)$ at $Q^2=4 GeV^2$ versus
$x$. Data are from ref.[2] and the solid line is the result of our fit.
The dashed line is the theoretical result after evolution at
$Q^2=10 GeV^2$.

\item[Fig.2a] The structure function $xF^{\nu N}_3(x)$ versus $x$. Data
are from ref.[13] at $Q^2=3 GeV^2$ and the solid line is the result of
our fit.

\item[Fig.2b] The antiquark contribution $x\bar q(x)=x\bar u(x)+ x\bar
d(x)+x\bar s(x)$ at $Q^2=3 GeV^2$ (full circles) and $Q^2=5 GeV^2$
(full triangles) versus $x$. Data are from ref.[12] and solid line is
the result of our fit.

\item[Fig.3] $xg^p_1(x)$ at $<Q^2>=10 GeV^2$ versus $x$. Data are
from ref.[5] (full squares) and ref.[6] (full circles) together with our
predictions at
$Q^2=10 GeV^2$. (Dotted line is the contribution of $\Delta u_{\rm
val}(x)$ only, dashed line is the contribution of $\Delta u(x)$ and
$\Delta \bar u(x)$, and solid line is the contribution of $\Delta
u(x)$, $\Delta \bar u(x)$ and $\Delta d_{\rm val}(x)$).

\item[Fig.4] $xg^n_1(x)$ at $<Q^2>=2 GeV^2$ versus $x$. Data are from
ref.[18] together with our predictions at $Q^2=2 GeV^2$ (Dashed line is
the contribution of $\Delta u(x)$ and $\Delta \bar u(x)$ only and solid
line is the contribution of $\Delta u(x)$, $\Delta \bar u(x)$ and
$\Delta d_{\rm val}(x)$).
\end{itemize}


\begin{thebibliography}{99}


\bibitem{1} R.D. Field and R.P. Feynman, \pr {\bf D15} (1977) 2590.

\bibitem{2} M. Arneodo \etal, (New Muon Collaboration), preprint
CERN-PPE/93-117 (revised version 22/02/94), to appear in Phys. Rev. D
and references therein.

\bibitem{3}K. Gottfried, \prl {\bf18} (1967) 1174.

\bibitem{4} T. Sloan, G. Smadja and R. Voss, \pr {\bf162} (1988) 45.

\bibitem{5} J. Ashman \etal, (European Muon Collaboration), \pl {\bf
B206}
(1988) 364; Nucl.\ Phys.\ {\bf B328} (1989) 1.

\bibitem{6} D. Adams \etal, (Spin Muon Collaboration), preprint
CERN-PPE/94-57.

\bibitem{7} Z. Dziembowski and J. Franklin, J.\ Phys.\ G: Nucl.
Part. Phys. {\bf 17} (1991) 213.

\bibitem{8} F. Buccella and J. Soffer, Mod.\ \pl A {\bf8} (1993) 225.

\bibitem{9} J. Babcock, E. Monsay and D. Sivers, Phys. Rev. {\bf D19}
(1979) 1483.

\bibitem{10} C. Bourrely, F. Buccella, G. Miele, G. Migliore, J. Soffer
and V. Tibullo, preprint CPT-93/P.2961, to appear in Z. Phys.

\bibitem{11} See for example J. Cleymans, I. Dadic and J. Joubert,
preprint BI-TP 92-45 (to appear in Z. Physik) and references therein.

\bibitem{12} C. Foudas \etal, \prl {\bf64} (1990) 1207; S.R. Mishra
\etal, \prl {\bf 68} (1992) 3499;
S.A. Rabinowitz \etal, \prl {\bf70} (1993) 134.

\bibitem{13} P.Z. Quintas \etal, (CCFR Collaboration) Phys. Rev. Lett.
{\bf 71} (1993) 1307; W.C. Leung \etal, (CCFR Collaboration) Phys.
Lett. {\bf B317} (1993) 655.

\bibitem{14} I. Abt \etal, ($H_1$ Collaboration), Nucl. Phys. {\bf
B407} (1993) 515. M. Derrick \etal, (ZEUS collaboration), Phys. Lett.
{\bf B316} (1993) 412.

\bibitem{15} G. Ballochi \etal, (UA6 Collaboration) Phys.
Lett. B {\bf 317} (1993) 250; P. Oberson \etal, talk at the 28$^{th}$
Rencontres de Moriond,  Les Arcs, March 1993; M.
Werlen \etal, talk at the International Europhysics Conference on HEP,
Marseille, July 1993.

\bibitem{16} M. Shaevitz \etal, (CCFR Collaboration) talk at Rencontres
de Physique de la Vall\'ee
d'Aoste, La Thuile, March 1993, preprint NEVIS R-1491.


\bibitem{17} A. Arneodo \etal, (New Muon Collaboration) Phys. Lett.
{\bf B309} (1993) 222.

\bibitem{18} P.L. Anthony \etal, (E142 Collaboration), \prl {\bf71}
(1993) 959.

\bibitem{19} B. Adeva \etal, (Spin Muon Collaboration), \pl {\bf B302}
(1993)
533.

\bibitem{20} J. Kwiecinski and D. Strozik-Kotlorz, Z. Phys. {\bf C48}
(1990) 315.

\bibitem{21} G. Altarelli and G. Parisi, Nucl. Phys. {\bf B126} (1977)
298.

\bibitem{22} P. Amaudruz \etal, (New Muon Collaboration) Nucl. Phys.
{\bf B371} (1992) 3.

\bibitem{23} J. Ellis and R.L. Jaffe, \pr {\bf D9} (1974) 1444.

\bibitem{24} G. Preparata and J. Soffer, \prl {\bf61} (1988) 1167;
{\bf62}
(1989) 1213 (E); G. Preparata, Ph.\ Ratcliffe and J. Soffer, \pl {\bf
B273} (1991) 306.

\bibitem{25} J.D. Bjorken, \pr {\bf148} (1966) 1467.

\bibitem{26} C. Bourrely and J. Soffer, preprint CPT-94/P.3000 (to
appear in Nucl. Phys. B) and references therein.

\end{thebibliography}
\end{document}